\documentclass[sigconf,twocolumn,anonymous=false]{acmart}
\setcopyright{none}
\renewcommand\footnotetextcopyrightpermission[1]{}

\AtBeginDocument{%
  }
\usepackage{listings}

\begin{document}
\title{Introducing Repository Stability}
\author{Giuseppe Destefanis$^1$, Silvia Bartolucci$^2$, Daniel Graziotin$^3$, Rumyana Neykova$^1$, Marco Ortu$^4$}
\affiliation{%
  \institution{$^1$Brunel University London, UK\\
  $^2$University College London, UK\\
  $^3$University of Hohenheim, Germany\\
  $^4$University of Cagliari, Italy}
  \country{}
  }
\email{{giuseppe.destefanis,rumyana.neykova}@brunel.ac.uk}
\email{s.bartolucci@ucl.ac.uk}
\email{graziotin@uni-hohenheim.de}
\email{marco.ortu@unica.it}
\renewcommand{\shortauthors}{Destefanis, et al.}
\begin{abstract}
Drawing from engineering systems and control theory, we introduce a framework to understand repository stability, which is a repository activity capacity to return to equilibrium following disturbances - such as a sudden influx of bug reports, key contributor departures, or a spike in feature requests. The framework quantifies stability through four indicators: commit patterns, issue resolution, pull request processing, and community engagement, measuring development consistency, problem-solving efficiency, integration effectiveness, and sustainable participation, respectively. These indicators are synthesized into a Composite Stability Index (CSI) that provides a normalized measure of repository health proxied by its stability. Finally, the framework introduces several important theoretical properties that validate its usefulness as a measure of repository health and stability. At a conceptual phase and open to debate, our work establishes mathematical criteria for evaluating repository stability and proposes new ways to understand sustainable development practices. The framework bridges control theory concepts with modern collaborative software development, providing a foundation for future empirical validation.
\end{abstract}


\maketitle

\section{Introduction}

Understanding when a software project is in a healthy state remains a critical yet unsolved challenge in software development. While repositories provide extensive data about project activities, from code changes to community interactions, current approaches struggle to convert this wealth of information into actionable insights about project health \cite{jansen2014measuring, xia2022predicting}. This gap affects both practitioners managing projects and researchers studying software development.

In this paper, we propose a new perspective: viewing project health through the lens of \textbf{stability} \cite{fayad2001thinking, salama2019stability}. We envision repositories as dynamic systems whose health manifests in their capacity to handle disturbances while maintaining consistent development practices.

Drawing from control theory \cite{filieri2015software}, we introduce a framework that reimagines repository analysis. Just as engineering systems naturally seek equilibrium after physical perturbations, we propose that healthy software projects demonstrate stability when facing disruptions - from sudden increases in bug reports to key contributor departures to spikes in feature requests. This perspective opens new possibilities for understanding and maintaining project health and sustainability \cite{venters2018software}.
Our framework improves repository analysis through four measurable indicators: commit frequency, issue resolution rate, pull request merge rate, and community activity, synthesized into a \textbf{Composite Stability Index (CSI)}. This mathematical foundation provides a new perspective in how we evaluate and predict project health, moving beyond traditional statistical approaches to capture the dynamic nature of software development.

The potential impact of our stability-based vision is threefold: (1) it enables systematic, quantitative evaluation of repository health, (2) it paves the way for data-driven project management through clear stability metrics, and (3) it opens new research directions in empirical software engineering. By bridging control theory and repository analysis, we offer a rigorous yet practical approach to a fundamental challenge.

We address the research question: \textit{How can we systematically define and measure repository stability to reflect and predict project sustainability?}

Our goal is to spark a paradigm shift in how we analyze and maintain software projects, establishing stability as a fundamental lens for understanding project health.

\section{Related Work}

The concept of stability has evolved across multiple domains, from control theory to software engineering \cite{filieri2015software}, yet a unified approach to repository stability remains elusive. While Lyapunov's work \cite{lyapunov1992general} and Leigh's contributions \cite{leigh2004control} established rigorous mathematical foundations for analyzing dynamic systems, their frameworks have not been adapted to the specific challenges of software repositories, which serve as the technical infrastructure where code and related artifacts of software projects are stored and managed.

The distributed systems community, through works like Tabuada et al. \cite{tabuada2012input} and Stankovic \cite{stankovic1985stability}, advanced our understanding of stability through bounded disturbances and performance metrics. However, these approaches, focused on technical states, have not captured the rich socio-technical dynamics that characterize modern repository evolution as part of broader project ecosystems.

Software engineering has explored stability from various perspectives. Jazayeri et al. \cite{jazayeri2002architectural} pioneered architectural stability measures for evolution, while Yau et al. \cite{yau1980some} developed foundational metrics for code stability. Yet these valuable approaches address only isolated aspects of repository dynamics without considering repositories as complete dynamic systems.

In parallel, significant research has addressed project health, which relates to but differs from repository stability. A project involves the full software development effort—people, processes, goals, and resources—while a repository is the technical artifact storing code and development history. The CHAOSS project represents a major initiative for standardizing health metrics for open source communities\footnote{\url{https://chaoss.community/}}. As Goggins et al. \cite{goggins2021open} note, open source project health fundamentally concerns "a project's ability to continue to produce quality software," but current approaches struggle to convert repository trace data into actionable insights about overall project health.

Project health has been conceptualized through various lenses: activity-based measures \cite{crowston2006}, community growth and diversity \cite{daniel2013}, and operational sustainability \cite{schweik2012}. Crowston et al. \cite{crowston2006} defined success through metrics related to project output, process quality, and team outcomes, while Daniel et al. \cite{daniel2013} emphasized the importance of social diversity in distributed communities.

Jansen \cite{jansen2014} expanded the scope beyond individual projects to ecosystem-level health through measures of productivity, robustness, and niche creation. The evaluation of project health often distinguishes between sustainability (the long-term ability to maintain development momentum) and survivability (resilience when facing disruptions). Raja and Tretter \cite{raja2012} introduced a viability index measuring vigor, resilience, and organization to assess a project's capacity to overcome challenges. Despite these advances, as Goggins et al. \cite{goggins2021open} argue, effective health assessment requires considering comparison, transparency, trajectory, and visualization principles when analyzing repository data.

Traditional repository mining, as demonstrated by Kagdi et al. \cite{kagdi2007survey} and Hammad et al. \cite{hammad2011automatically}, has revealed valuable historical patterns \cite{synovic2022snapshot}. However, these approaches often treat repositories as static artifacts, missing the opportunity to understand their dynamic, evolving nature as systems that exhibit stability properties analogous to those in control theory.

Salama et al. \cite{salama2019stability} surveyed stability across software artifacts; their work revealed an important gap: the absence of a unified theoretical foundation for repository stability. Our work aims at filling this gap by integrating control theory principles with repository dynamics, enabling systematic analysis of both structural and behavioral stability through well-defined metrics and thresholds. 

While existing frameworks emphasize narratives and contextual understanding across multiple dimensions of project health, our stability-based approach offers a complementary mathematical perspective by applying control theory principles to quantify a repository's ability to maintain equilibrium after experiencing disturbances. This foundation enables systematic evaluation of stability as a key dimension of overall repository health, bridging theoretical rigor with practical utility.

\section{Theoretical Foundation from Control Theory}
In control theory, stability is fundamentally concerned with a system's behavior over time and its response to perturbations. A system is considered stable if, when disturbed from equilibrium, it tends to return to its equilibrium state. More formally, for a dynamical system described by

\begin{equation}
    \dot{x} = f(x,t)\ ,
\end{equation}
where $x$ represents the state vector and $t$ represents time, stability is often characterized using Lyapunov stability theory. A system is considered stable if for any $\epsilon > 0$, there exists a $\delta > 0$ such that:

\begin{equation}
    \|x(t_0)\| < \delta \implies \|x(t)\| < \epsilon, ~\forall t \geq t_0 \ .
\end{equation}

\subsection{Repository as a Dynamical System}

The conceptualization of a repository as a dynamical system emerges from key observations about software development patterns. Software repositories exhibit continuous evolution through time, with state changes driven by developer interactions, mirroring classical dynamical systems in physics or engineering. These repositories feature complex feedback mechanisms where code changes trigger reviews and subsequent modifications, creating interconnected activity cycles. They encompass both deterministic elements, such as automated workflows, and stochastic components such as varying developer activity patterns, resembling mixed deterministic-stochastic systems.
Repositories demonstrate equilibrium-seeking behavior, alternating between periods of intense development and stabilization, analogous to classical dynamical systems. The measurable metrics --commits, issues, pull requests, and branch activities -- serve as state variables that evolve according to rules governed by technical and social factors. This conceptualization enables the application of dynamical systems analysis techniques to quantify and understand repository stability.

Let $R(t)$ represent the state of a repository at time $t$, defined as a vector:

\begin{equation}
    R(t) = [c(t), i(t), p(t), a(t)]^T \ ,
\end{equation}

where:
\begin{itemize}
    \item $c(t)$: Commit frequency function;
    \item $i(t)$: Issue resolution rate function;
    \item $p(t)$: Pull request merge rate function;
    \item $a(t)$: Activity engagement function.
\end{itemize}

These four components have been specifically chosen based on available repository metrics that represent fundamental dimensions of repository activity and health. The commit frequency function $c(t)$ captures the development momentum and intensity of code changes, providing insights into the project's active development patterns through commit timestamps and frequencies. The issue resolution rate function $i(t)$ reflects the project's ability to handle and resolve problems, calculated through the analysis of issue creation and closure timestamps. The pull request merge rate function $p(t)$ measures the effectiveness of the code review and integration processes, derived from pull request lifecycle data. Finally, the activity engagement function $a(t)$ provides insight into the overall repository engagement through comment activity and interaction patterns, replacing the branch lifetime metric with a more readily available measure of repository vitality.
Each component is defined as follows:

\subsubsection*{Commit Frequency Function}
\begin{equation}
    c(t) = \frac{N_c(t, t+\Delta t)}{\Delta t} \ ,
\end{equation}
where $N_c(t, t+\Delta t)$ represents the number of commits in the time interval $[t, t+\Delta t]$, directly measurable from our commit history data.

\subsubsection*{Issue Resolution Rate}
\begin{equation}
    i(t) = \frac{N_i^{closed}(t, t+\Delta t)}{N_i^{total}(t)} \cdot \frac{1}{1 + \overline{T}_{resolution}(t)} \ ,
\end{equation}
where $N_i^{closed}$ represents closed issues, $N_i^{total}$ represents total issues, and $\overline{T}_{resolution}(t)$ is the average resolution time for issues closed in the interval $[t, t+\Delta t]$.

\subsubsection*{Pull Request Merge Rate}
\begin{equation}
    p(t) = \frac{N_p^{merged}(t, t+\Delta t)}{N_p^{total}(t)} \cdot \frac{1}{1 + \overline{T}_{review}(t)} \ ,
\end{equation}
where $N_p^{merged}$ represents merged pull requests, $N_p^{total}$ represents total pull requests, and $\overline{T}_{review}(t)$ is the average review time for pull requests in the interval $[t, t+\Delta t]$.

\subsubsection*{Activity Engagement Function}
\label{a(t)}
\begin{equation}
    a(t) = \frac{N_{comments}(t, t+\Delta t)}{N_{issues}(t) + N_{prs}(t)} \cdot \frac{N_{active\_users}(t, t+\Delta t)}{N_{total\_users}(t)}
    \label{eq:engagemnt}
\end{equation}

where $N_{comments}(t, t+\Delta t)$ represents the number of comments in the interval $[t, t+\Delta t]$, $N_{issues}(t) = N_i^{total}(t) - N_i^{closed}(t)$ and $N_{prs}(t) = N_p^{total}(t) - N_p^{merged}(t)$ represent the number of open issues and pull requests at time $t$ respectively, and $N_{active\_users}(t, t+\Delta t)$ represents users who have engaged with the repository through comments, commits, or pull requests in the interval $[t, t+\Delta t]$.

\section{Repository-Stability Definition}
We define the stability of a repository through a framework that considers both the individual metrics and their interrelationships. A repository's stability is characterized by consistent development patterns, efficient issue resolution, effective pull request management, and sustained community engagement.
\begin{definition}[Stability]
A repository $R(t)$ is considered stable if it satisfies all of the following criteria over an observation period $[t_0, t_0 + T]$. \end{definition}\vspace{0.1cm}
1. \textbf{Commit Pattern Stability:}
\begin{equation}
\left|\frac{dc(t)}{dt}\right| \leq \alpha_c, ~~\forall t \in [t_0, t_0 + T] \ ,
\end{equation}
where $\alpha_c$ represents the maximum allowable rate of change in commit frequency (the highest acceptable value that the rate of change in commit frequency can have for a repository to still be considered stable). This criterion ensures that development activity maintains a consistent rhythm without extreme fluctuations that could indicate project's instability. The threshold is defined as
\begin{equation}
\alpha_c = \frac{\sigma_{\text{daily commits}}}{\mu_{\text{daily commits}}} \leq 0.5 \ .
\end{equation}
The coefficient of variation threshold $\alpha_c$ provides a measure of commit frequency stability. We propose $\alpha_c = 0.5$ as an initial value because this would indicate that the standard deviation remains less than half the mean, thus allowing for natural variations, while potentially identifying problematic patterns such as long periods of inactivity followed by a burst of commits. This relative measure accommodates projects of different sizes and activity levels, although empirical validation is needed to confirm its effectiveness.
\vspace{0.1cm}\\
2. \textbf{Issue Management Stability:}
\begin{equation}
   i(t) \geq \beta_i ~\text{ and } ~\overline{T}_{resolution}(t) \leq \tau_i, ~\forall t \in [t_0, t_0 + T] \ ,
\end{equation}
where
\begin{itemize}
   \item $\beta_i$ is the minimum acceptable issue resolution rate (proposed value: 0.3);
   \item $\tau_i$ is the maximum acceptable average resolution time (proposed value: 14 days);
   \item $\overline{T}_{resolution}(t)$ is the moving average of issue resolution time.
\end{itemize}
The thresholds for issue management stability reflect considerations in modern software development practices. We propose $\beta_i=0.3$ for the minimum resolution rate, acknowledging that not all issues require immediate resolution -- some may be feature requests, duplicates, or issues that soon become obsolete. For the maximum resolution time, here we consider $14$ days, aligning with sprint cycles of two weeks in agile development. 
These are initial proposed values, but empirical validation is required to determine their effectiveness across different project sizes and development intensities. \vspace{0.1cm}\\
3. \textbf{Pull Request Processing Stability:}
\begin{equation}
   p(t) \geq \beta_p ~\text{ and } ~\overline{T}_{review}(t) \leq \tau_p, ~\forall t \in [t_0, t_0 + T] \ ,
\end{equation}
where
\begin{itemize}
   \item $\beta_p$ is the minimum acceptable pull request merge rate (threshold: 0.4);
   \item $\tau_p$ is the maximum acceptable average review time (threshold: 5 days);
   \item $\overline{T}_{review}(t)$ is the moving average of pull request review time.
\end{itemize}

The proposed $\beta_p=0.4$ pull request merge rate acknowledges that, while some requests warrant rejection due to experimental nature or needed revisions, maintaining an overly low rate may unnecessarily impede contributions. The five-day maximum review period strikes a balance between enabling thorough code evaluation and maintaining development momentum. These initial thresholds aim to optimize between quality control and velocity, though their effectiveness will need to be validated through implementation data.
\vspace{0.1cm}\\
4. \textbf{Community Engagement Stability:}
\begin{equation}
   a(t) \geq \gamma_a ~\text{ and } ~\frac{N_{active\_users}(t)}{N_{total\_users}(t)} \geq \delta_a, ~\forall t \in [t_0, t_0 + T] \ ,
\end{equation}
where
\begin{itemize}
   \item $\gamma_a$ is the minimum acceptable activity ratio (threshold: proposed value 0.25);
   \item $\delta_a$ is the minimum acceptable active user ratio (threshold: proposed value 0.15).
\end{itemize}

The activity function $a(t)$ here is the same as defined in Section \ref{a(t)} (see Eq. \ref{eq:engagemnt}), measuring both the intensity of interactions through comment ratios and the breadth of community participation through user activity ratios. The thresholds ensure that the repository maintains both sufficient discussion density, relative to open items, and broad community involvement.

\subsection{Stability Thresholds}
We represent the proposed threshold values for each metric as a threshold matrix:
\begin{equation}
   \text{T. Matrix } \Theta = \begin{bmatrix} 
   \alpha_c & \beta_i & \tau_i & \beta_p & \tau_p & \gamma_a & \delta_a \\
   0.5 & 0.3 & 14 & 0.4 & 5 & 0.25 & 0.15
   \end{bmatrix} \ .
\end{equation}
These initial threshold values represent educated estimations based on the authors' experience with repository analysis and software development processes. The commit pattern threshold ($\alpha_c = 0.5$) allows for natural variations while still identifying erratic behavior. Issue and pull request thresholds balance timely response with thorough processing. Community engagement thresholds aim to ensure broad participation. We acknowledge these are initial proposals that must be empirically validated in future studies.

\subsection{Composite Stability Index}
To provide a single quantitative measure of stability, we define the Composite Stability Index (CSI) as a weighted sum of normalized stability metrics:

\begin{equation}
   CSI(t) = w_c\phi_c(c(t)) + w_i\phi_i(i(t)) + w_p\phi_p(p(t)) + w_a\phi_a(a(t)) \ .
\end{equation}

The weights reflect the relative importance of each stability component, and their sum must equal $1$, with commit pattern stability given slightly higher priority due to its direct reflection of development activity. The weight vector is defined as

\begin{equation}
    W = [w_c, w_i, w_p, w_a] = [0.3, 0.25, 0.25, 0.2] \ .
\end{equation}

Each component is normalized through a function $\phi_k$ that maps the raw metrics to a $[0,1]$ scale, ensuring comparable contributions to the final index:

\begin{equation}
    \phi_k(x) = \begin{cases}
    1 - \frac{|x - \mu_k|}{\sigma_k} & \text{if } ~|x - \mu_k| \leq \ \sigma_k \ , \\
    0 & \text{otherwise} \ .
    \end{cases}
\end{equation}

The target values $\mu_k$ and acceptable deviations $\sigma_k$ are defined for each component based on their respective thresholds:

\begin{itemize}
   \item For commit pattern stability ($\phi_c$):
   \begin{itemize}
      \item $\mu_c = 0.25$ (target coefficient of variation);
      \item $\sigma_c = 0.25$ (allowing variation up to the 0.5 threshold).
   \end{itemize}
   
   \item For issue management stability ($\phi_i$):
   \begin{itemize}
      \item $\mu_i = 0.4$ (target resolution rate);
      \item $\sigma_i = 0.1$ (allowing variation down to the 0.3 threshold).
   \end{itemize}
   
   \item For pull request stability ($\phi_p$):
   \begin{itemize}
      \item $\mu_p = 0.5$ (target merge rate);
      \item $\sigma_p = 0.1$ (allowing variation down to the 0.4 threshold).
   \end{itemize}
   
   \item For community engagement stability ($\phi_a$):
   \begin{itemize}
      \item $\mu_a = 0.35$ (target activity ratio);
      \item $\sigma_a = 0.1$ (allowing variation down to the 0.25 threshold).
   \end{itemize}
\end{itemize}

Target values ($\mu_k$) and deviations ($\sigma_k$) are set to encourage improvement beyond minimum thresholds, while ensuring that meeting these minimums contributes positively to the CSI. We propose a CSI threshold of $0.7$ for overall stability, requiring components to perform consistently above their minimum thresholds, while accommodating temporary fluctuations. This weighted and normalized scoring system enables both monitoring overall repository health and diagnosing specific areas requiring attention when the CSI falls below the threshold.

\section{Mathematical Properties}
The proposed Repository-Stability framework exhibits important theoretical properties that validate its usefulness as a measure of repository health and stability:

1. \textbf{Boundedness}
The Composite Stability Index (CSI) is bounded by design:
\begin{equation}
   0 \leq CSI(t) \leq 1, \forall t \in [t_0, t_0 + T] \ .
\end{equation}
This boundedness property ensures that stability measures are normalized and comparable across different repositories and time periods.

2. \textbf{Piecewise Continuity}
The stability measure should be piecewise continuous with respect to all its component metrics. Within each piece defined by $|x - \mu_k| \leq \sigma_k$, for any metrics $m_1, m_2$:
\begin{equation}
   |m_1 - m_2| < \delta \implies |CSI(m_1) - CSI(m_2)| < \epsilon \ .
\end{equation}
This property ensures that small changes in repository metrics result in proportional changes in the stability measure within each regime.

3. \textbf{Long-Term Stability} If a repository's metrics stabilize over time (i.e., the system approaches an equilibrium), the Composite Stability Index (CSI) should converge to a constant value:
\begin{equation}
    \lim_{t \to \infty} CSI(t) = \text{constant} \ .
\end{equation}

This property reflects the long-term stability of a repository. In engineering, stable systems often approach equilibrium states as disturbances diminish over time. For a repository, this could correspond to regular commit patterns, balanced issue resolution and pull request processing, and consistent community engagement.

If each metric $\phi_k(x_k(t))$ stabilizes over time, such that:
\begin{equation}
    \lim_{t \to \infty} \phi_k(x_k(t)) = \phi_k^\infty \ ,
\end{equation}
then the Composite Stability Index (CSI) converges to:
\begin{equation}
    \lim_{t \to \infty} CSI(t) = \sum_{k} w_k \phi_k^\infty\ ,
\end{equation}
where $w_k$ are the weights assigned to each metric. This result shows that the CSI becomes a weighted sum of the asymptotic values of the individual metrics, providing a clear mathematical representation of the long-term behavior of the repository.

\textbf{Conditions for Convergence:} Metrics such as commit frequency, issue resolution rate, and pull request merge rate must exhibit bounded fluctuations or a decaying trend toward equilibrium. For instance, commit frequency $c(t)$ stabilizes if:
\begin{equation}
    \frac{dc(t)}{dt} \to 0 \quad \text{as} \quad t \to \infty \ .
\end{equation}

While theoretical at this stage, this property could be empirically tested by analyzing historical data from repositories over long periods. Stable repositories should display convergence trends in their CSI values, whereas repositories with irregular or unsustainable practices may show divergence or high variability.

\textbf{Practical Implications:} 
Repositories with erratic behavior (e.g., bursts of activity followed by dormancy) may not exhibit convergence, signaling instability, while stable repositories demonstrate convergence, reflecting healthy long-term development practices. This property provides a diagnostic tool for identifying repositories requiring intervention to achieve sustained stability. When stability metrics consistently fall below their thresholds, automated monitoring systems can identify specific areas requiring attention. For commit pattern instability, the system would highlight development cycle issues; for degrading pull request processing, it would indicate review process inefficiencies. Developers can then implement targeted interventions, which may include restructuring sprint planning to stabilize commit patterns, or redistributing review responsibilities to improve pull request processing. These metric-specific corrections allow teams to address stability issues before they compromise overall project health, transforming theoretical stability measures into practical maintenance tools within existing development workflows.

\section{Conclusion and Open Challenges}
We introduced a theoretical framework for analyzing repository stability through the lens of control theory. The framework's four core components --- commit patterns, issue resolution, pull request processing, and community engagement --- provide a complete view of repository stability, while remaining computationally tractable. Through the newly introduced Composite Stability Index, we offer a tool to measure and monitor repository health through its stability characteristics. Our framework makes several fundamental assumptions: that repositories exhibit equilibrium-seeking behavior similar to physical systems, that stability can be meaningfully quantified through our chosen metrics, and that these measures capture the essential dynamics of software development. These assumptions, while grounded in software engineering practices, open critical questions for the research community:\\
\textit{1) To what extent can universal stability thresholds be established, or do they inherently depend on repository context?}
This question builds on prior work by Jansen \cite{jansen2014}, who observed that metrics may have different meanings for different projects. While our framework proposes initial thresholds (e.g., $\alpha_c = 0.5$ for commit pattern stability), empirical research across diverse repositories is needed to determine whether universal thresholds exist or if contextual adaptation is necessary. A promising hypothesis is that repositories within similar domains (e.g., system libraries vs. web applications) share similar optimal stability thresholds.\\
\textit{2) Under what conditions does the control theory analogy hold for software repositories, and which differences could be brought by open-source and proprietary software development?} This question extends Salama et al.'s \cite{salama2019stability} work on stability concepts in software engineering, exploring whether open-source repositories follow different equilibrium patterns than proprietary ones. Factors such as governance structures \cite{raja2012} and corporate engagement \cite{daniel2013} likely influence stability dynamics. We hypothesize that repositories with corporate backing exhibit different recovery patterns following disturbances than community-driven projects.\\
\textit{3) What mathematical models could better represent the complex interactions between stability components while maintaining practical applicability?} Building on Filieri et al.'s \cite{filieri2015software} application of control theory to software engineering, future work could explore non-linear models that capture emergent interactions between stability components. For instance, the relationship between issue resolution and community engagement might be better modeled through coupled differential equations that capture feedback mechanisms.\\
\textit{4) How do different socio-technical factors --- from team structure to development methodologies to community norms --- affect our notion of stability?} This question connects with Goggins et al.'s \cite{goggins2021open} emphasis on social context in health assessment. We hypothesize that repositories with explicit governance mechanisms (e.g., defined contribution processes) exhibit higher stability in the face of similar disturbances compared to those with ad-hoc structures. Future research could categorize repositories based on these socio-technical factors and analyze their CSI patterns.\\
\textit{5) How can this theoretical framework inform practical tools for project maintenance while accounting for the social nature of software development?} This bridges theory with practice. Stability metrics could be integrated into project dashboards that provide early warnings of declining stability, with visualizations that will help interpret complex stability data in context.
These questions point to rich opportunities in both theoretical and empirical directions --- from validating the framework through large-scale repository analysis, to empirically defining threshold values, to investigating non-linear stability metrics --- potentially transforming how we understand and evaluate software projects' health. 

Future work should directly compare our stability framework with traditional repository health metrics to validate whether our approach captures dynamic aspects of development that static metrics miss. User studies with development teams could reveal how our framework translates from a theoretical model to practical monitoring tools.

\bibliographystyle{unsrt}

\end{document}